\documentclass[11pt,aps]{revtex4}

\usepackage{epsfig,amsmath,amsfonts}
\usepackage{times}

\def\>{\rangle}
\def\<{\langle}

\newcommand{\ignore}[1]{}

\def\bmat{\left[\begin{matrix}} 
\def\emat{\end{matrix}\right]}

\newcommand{\proj}[1]{\ket{#1}\bra{#1}}
\newcommand{\bra}[1]{\langle#1|}
\newcommand{\ket}[1]{{|}#1{\rangle}}

\newcommand{\gt}[1]{\textsc{#1}}
\newcommand{\cnot}{\gt{CNOT}}
\newcommand{\meas}{\gt{Measure}}

\newcommand{\eccode}[3]{$[\![#1,#2,#3]\!]\;$}
\newcommand{\steane}{\eccode{7}{1}{3}}

\begin{document}

\title{Resource Requirements for Fault-Tolerant Quantum Simulation:
The Transverse Ising Model Ground State}

\author{Craig R.\ Clark$^1$}
\author{Tzvetan S.\ Metodi$^2$}
\author{Samuel D.\ Gasster$^2$}
\author{Kenneth R.\ Brown$^1$}
\email{ken.brown@chemistry.gatech.edu}
\affiliation{$^1$School of Chemistry and Biochemistry and Division of 
Computational Science and Engineering,\\ 
Georgia Institute of Technology, Atlanta, GA 30332-0400}
\affiliation{$^2$Computer Systems Research Department, 
The Aerospace Corporation, El Segundo, CA 90245-4691}

\date{\today}

\begin{abstract}

We estimate the resource requirements, the total number of physical qubits and
computational time, required to compute the ground state energy of a 1-D
quantum Transverse Ising Model (TIM) of $N$ spin-1/2 particles, as a function
of the system size and the numerical precision. This estimate is based on
analyzing the impact of fault-tolerant quantum error correction in the context
of the Quantum Logic Array (QLA) architecture. Our results show that due to
the exponential scaling of the computational time with the desired precision
of the energy, significant amount of error correciton is required to implement
the TIM problem. Comparison of our results to the resource requirements for
a fault-tolerant implementation of Shor's quantum factoring algorithm reveals that
the required logical qubit reliability is similar for both the TIM problem and
the factoring problem.

\end{abstract}


\maketitle


\section{Introduction \label{sec:intro}}

The calculation of the basic properties of quantum systems (eigenstates and
eigenvalues) remains a challenging problem for computational science.  One of
the most significant issues is the exponential scaling of the computational
resource requirements with the number of particles and degrees of freedom,
which for even a small number of particles ($\sim100$) exceeds the
capabilities of current computer systems.  In 1982 Feynman addressed this
problem by proposing that it may be possible to use one quantum system as the
basis for the simulation of another \cite{Feynman1982}. This was the early
promise of quantum simulation, and one of the original motivations for quantum
computing. Since that time, many researchers have investigated different
approaches to quantum simulation
\cite{Lloyd1996,Abrams1997,Abrams1999,Zalka1998,Aspuru-Guzik2005,Kassal:08}.
For example, Abrams and Lloyd  have proposed a quantum algorithm for the
efficient computation of eigenvalues and eigenvectors using a quantum computer
\cite{Abrams1999}. Many of the investigations into quantum simulation have
assumed ideal performance from the underlying components resulting in
optimistic estimates for the quantum computer resource requirements (number of
qubits and time to completion). It is well known, however, that in order to
address the effects of decoherence and other sources of faults and errors in
the implementation of qubits and gates it is necessary to incorporate
fault-tolerant quantum error correction into an estimate of the resource
requirements.

In this paper we estimate the resource requirements for a quantum simulation
of the ground state energy for the 1-D quantum Transverse Ising Model,
specifically incorporating the impact of fault-tolerant quantum error
correction. We apply the general approach of Abrams and Lloyd
\cite{Abrams1997,Abrams1999}, and compute estimates for the total number of
physical qubits and computational time as a function of the number of
particles (N) and required numerical precision (M) in the estimate of the
ground state energy.

We have chosen to study the resource requirements for computing the ground
state energy for the 1-D quantum TIM since this model is well studied in the
literature and has an analytical solution
\cite{Pfeuty:1970rm,Sachdev:1999hl,Jouzapavicius1997}. The relevant details of the TIM are
summarized in Section~\ref{sec:ising}. In Section~\ref{sec:qsim}, we map the
calculation of the TIM ground state energy onto a quantum phase estimation
circuit that includes the effects of fault-tolerant quantum error correction.
The required unitary transformations are decomposed into one qubit gates and
two-qubit controlled-not gates using gate identities and the Trotter formula.
The one-qubit gates are approximated by a set of gates which can be executed
fault-tolerantly using the Solovay-Kitaev theorem \cite{KSV}. In
Section~\ref{sec:faultTolerance}, the quantum circuit is mapped onto the
Quantum Logic Array (QLA) architecture model, previously described by Metodi,
et al. \cite{Metodi:05}. Our final results, utilizing the QLA architecture,
are given in Section \ref{sec:oneD} and a discussion of how improving the
state of the art in the underlying technology affects the performance for
executing the TIM problem. In Section \ref{sec:higherDim}, we extend our
resource estimate from 1-D to higher dimensions. Since the QLA architecture
was developed to study the fault-tolerant resource requirements for Shor's
quantum factoring algorithm \cite{Shor:1994qy}, we compare our present results
for the TIM quantum simulation with previous analysis of the the resource
requirements for Shor's algorithm, in Section V. Finally, our conculsions are
presented in Section ~\ref{sec:conclusion}.


%
%
\section{Transverse Ising Model\label{sec:ising}}

The 1-D Transverse Ising Model is one of the simplest models exhibiting a
quantum phase transition at zero temperature
\cite{Pfeuty:1970rm,Elliott:1970vn,Jullien:1978yg,Sachdev:1999hl}. The
calculation of the ground state energy of the TIM varies from analytically
solvable in the linear case \cite{Pfeuty:1970rm} to computationally
inneficient for frustrated 2-D lattices \cite{Moessner2003}. For example, the
calculation of the magnetic behavior of frustrated Ising antiferromagnets
requires computationally intensive Monte-Carlo simulations \cite{Hu2008}.
Given the difficulty of the generic problem and the centrality of the TIM to
studies of quantum phase transitions and quantum annealing, the TIM is a good
benchmark model for quantum computation studies.

The 1-D Transverse Ising Model consists of $N$-spin-1/2 particles at each
site of a one dimmensional linear lattice (with the spin axis along the $z$-axis) 
in an external magnetic field along to 
$x$-axis. The Hamiltonian for this system, $H_I$, may be written as
\cite{Sachdev:1999hl}:
\begin{equation}\label{eqn:IsingHamiltonian}
H_I =  \sum_i \Gamma \sigma^x_i +  \sum_{\langle i,j\rangle} J_{ij}\sigma^z_i \sigma^z_j ,
\end{equation}
where $J$ is the spin-spin interaction energy, $\Gamma$ is the
coupling constant and related to the strength of the external magnetic field
along the $\hat{x}$-direction, and $\langle i,j\rangle$ implies a sum only
over nearest-neighbors. $\sigma^x_i$ and $\sigma^z_i$ are the Pauli spin
operators for the $i$th spin, and we set $\hbar = 1$ throughout this paper.

In present work we focus is on the 1-D linear chain TIM of N-spins with
constant Ising interaction energy $J_{ij} = -J$. The ground state of the
system is determined by the ratio of $g =\Gamma/J$. For the large magnetic
field case, $g >> 1$ the system is paramagnetic with all the spins aligned
along the $\hat{x}$ axis, and in the limit of small magnetic field, $g << 1$,
the system has two degenerate ferromagnetic ground states, parallel and
anti-parallel to the $\hat{z}$ axis.  In the intermediate range of magnetic
field strength the linear 1-D TIM exhibits a quantum phase transition at $g =
1$ \cite{Sachdev:1999hl}.

The TIM Hamiltonian in Equation~\ref{eqn:IsingHamiltonian}, for the 1-D case
with constant coupling can be rewritten as:
\begin{equation}\label{eqn:IsingHamiltonian2} 
H_I =  - J \left (\sum_{j=1}^{N}  g X_j +  \sum_{j=1}^{N-1} Z_jZ_{j+1}  \right ) 
\end{equation}
where the Pauli spin operators are replaced with their corresponding matrix
operators $X_j, Z_j$.  For the 1-D TIM, the ground state energy can be
calculated analytically in the limit of large N\cite{Pfeuty:1970rm}.  In the case of a finite
number of spins with non-uniform spin-spin interactions ($J$ not constant), it
is possible to efficiently simulate the TIM using either the  Monte-Carlo
method \cite{Santos2000} or the density matrix renormalization group approach
\cite{Jouzapavicius1997}. The challenge for classical computers comes from the
2-D TIM on a frustrated lattice where the simulation scales exponentially with
$N$. Applying the quantum phase estimation circuit to
calculate the ground state energy of the TIM requires physical qubit
resources, which scale polynomially with $N$, and the number of computational
time steps is also polynomial in $N$. In addition, just as the
complexity of the problem is independent of the lattice dimension and layout
when applying classical brute force diagonalization, the amount of resources
required to apply the quantum phase estimation circuit is largely independent
of the dimensionality of the TIM Hamiltonian.

\section{TIM Quantum Simulation Resource Estimates \label{sec:qsim}}

Our approach to estimating the resource requirements for the TIM ground-state
energy calculation with Hamiltonian $H_I$ involves two steps. First, we follow
the approach of Abrams and Lloyd and map the problem of computing the
eigenvalues of the TIM Hamiltonian in Equation~\ref{eqn:IsingHamiltonian2}
onto a phase estimation quantum circuit \cite{Abrams1997,Abrams1999}. Second,
we decompose each operation in the phase estimation circuit into a set of
universal gates that can be implemented fault-tolerantly within the context of
the QLA architecture. This allows us an
accurate estimate of the resources in a fault-tolerant environment.


\subsection{Phase estimation circuit \label{sec:peca} }

The phase estimation algorithm allows one to calculate an $M$-bit estimate of
the phase $\phi$ of the eigenvalue $e^{\displaystyle -i2\pi\phi}$ of the time
evolution unitary operator $U(\tau) = e^{\displaystyle -iH_I\tau}$, where the
time $\tau$ is constant throughout the implementation of the phase estimation
algorithm. The desired energy eigenvalue $E$ of $H_I$ can be computed using
$\phi$ by calculating $E = \frac{2\pi\phi}{\tau}$.

The value of $\tau$ is determined by the fact that the output from the phase
estimation algorithm is the binary fraction $0.x_1~...~x_M$, which is less
than one \cite{Abrams1997,Abrams1999}. In order to ensure that this result is
a valid approximation of the phase $\phi$, we must set the parameter $\tau$
such that $\tau < 2\pi/E$, which corresponds to $\phi < 1$. For the 1-D TIM,
the magnitude of the ground-state energy $|E_g|$ is bounded by $NJ (1 + g)$
\cite{Pfeuty:1970rm}. In the region near the phase transition $g\approx 1$, 
we choose $\tau$=$(10 J N)^{-1}$, which satisfies $\tau<2\pi/|E_g|$.

\begin{figure}
\centering
\setlength{\unitlength}{1mm}
\includegraphics[width=1.0\textwidth]{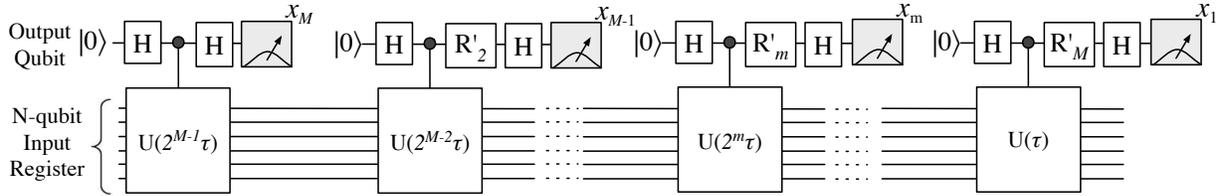}
\caption{The circuit for implementing the phase estimation algorithm using one 
continuously recycled control qubit.}\label{fig:OneControlQubit}
\end{figure}

The quantum circuit for implementing the phase estimation algorithm is shown
in Figure~\ref{fig:OneControlQubit}. The circuit consists of two quantum registers:
an $N$-qubit input quantum register prepared in an initial quantum state
$\ket{\Psi}$, and an output quantum register consisting of a single
qubit recycled $M$ times \cite{Parker.S:2000aa,Cleve1997}. Each of the $N$
qubits in the input register corresponds to one of the $N$ spin-$1/2$
particles in the TIM model \cite{Brown:2006qf}. At the beggining of each of the
$M$ steps in the algorithm, the output qubit is prepared into the state
$\frac{1}{\sqrt{2}}(\ket{0}+\ket{1})$ using a Hadamard (H) gate. The H gate is
followed by a controlled power of $U(\tau)$, denoted with $U(2^m\tau)$,
applied on the input register, where $0 \leq m \leq M-1$.

Letting $j$ denote to the $j$th step in the circuit, each time the output
qubit is measured (meter symbols) the result is in the $m$th bit in the estimate
of $\phi$, following the rotation of the output qubit via the gate:
\begin{equation} 
R_{j}=\proj{0}+\exp \left( i\pi \sum_{m=M+2-j}^M  \frac{2^{M+1}x_m}{2^{m+j}}\right )\proj{1}
\end{equation} 
where the gate $R_j$ corresponds to the application of the Quantum Fourier
Transform on the output qubit at each step \cite{Parker.S:2000aa,Cleve1997}.
The result after each of the $M$ measurements is an $M$-bit binary string
$\{x_1x_2\ldots x_{M}\}$, which corresponds the $M$-bit approximation of
$\phi$ given by $0.x_1~...~x_M$. Using this estimate of $\phi$, the
corresponding energy eigenvalue $E = \displaystyle\frac{2\pi\phi}{\tau}$ will
be the ground-state energy $E_g$ with probability
equal to $|\langle\Psi{|}\Psi_g\rangle|^2$ \cite{Abrams1997}, where
$\ket{\Psi_g}$ is the ground eigenstate of $H_I$. 

To maximize the probability of success $|\langle\Psi{|}\Psi_g\rangle|^2$, 
the initial quantum state $\ket{\Psi}$
should be an approximation of the ground state $\ket{\Psi_g}$. For arbitrary
Hamiltonians the preparation of an approximation to $\ket{\Psi_g}$ is
generally computationally difficult \cite{Kempe2006,Biamonte2008}. For certain
cases, the preparation can be accomplished using classical approximation
techniques to calculate an estimated wavefunction or adiabatic quantum state
preparation techniques \cite{Brown:2006qf,Aspuru-Guzik2005}. If the state can
be prepared adiabatically,  the resource requirements for preparing
$\ket{\Psi}$ are comparable in complexity to the resource requirements for
implementing the circuit for the phase estimation algorithm shown in Figure
\ref{fig:OneControlQubit} \cite{Brown:2006qf}. 
For this reason, we focus our analysis on estimating
the number of computational time steps and qubits required to implement the
circuit, assuming that the input register has been already prepared into the
$N$-qubit quantum state $\ket{\Psi}$.

\subsection{Decomposition of the TIM quantum circuit into fault-tolerant gates \label{sec:dftg}}

Figure \ref{fig:OneControlQubit} in Section \ref{sec:peca} shows the TIM
circuit at a high-level, involving $N+1$ unitary operators. In
this section, each unitary operation of the circuit is decomposed into a set
of basic one and two qubit gates which can be implemented fault-tolerantly
using the QLA architecture. The set of basic gates used is
\begin{equation}\label{eqn:basicgates}
\{X,Z,H,T,S,\cnot,\meas\}
\end{equation}
where \meas~ is a single qubit measurement in the $\hat{z}$ basis, \cnot~
denotes the two-qubit controlled-NOT gate, and $T$ and $S$ gates are
single-qubit rotations around the $\hat{z}$-axis by $\pi/4$ and $\pi/2$
radians respectively. The high-level circuit operations which require
decomposition
are the controlled-$U(2^m\tau)$ gates and each $R_j$ gate. 

The Controlled-$U(2^m\tau)$ gate can be decomposed using the second-order
Trotter formula \cite{Suzuki:92,Nielsen:2001fr}. First, $H_I$ is broken into
two terms: $H_X=\sum_{j=0}^{N} gX_j$, representing the transverse magnetic
field, and $H_{ZZ}=\sum_{j=0}^{N-1}Z_jZ_{j+1}$, representing the Ising
interactions. By considering the related unitary operators 
\begin{eqnarray}\label{eqn:UTrotter3}
\label{eqn:ux} U_x(2\tau)  & = & \displaystyle\prod_{j=1}^{N} \exp(-ig\tau X_j) \\ 
\label{eqn:uzz} U_{zz}(2\tau)    & = & \displaystyle\prod_{j=1}^{N-1} \exp(-i\tau Z_{j}Z_{j+1}),
\end{eqnarray}
where we set $g=1$, as discussed in Section \ref{sec:ising}. We can
construct the Totter approximation of $U(2^m\tau)$, denoted by
$\tilde{U}(2^m\tau)$ as:
\begin{eqnarray}\label{eqn:UTrotter}
\nonumber U(2^m\tau) & = &\left[U_x(\theta)~U_{zz}(2\theta)~U_x(\theta)\right]^k + \epsilon_T\\
					 & = & \tilde{U}(2^m\tau)  + \epsilon_T,
\end{eqnarray}
where $\theta = (2^m\tau/k)$ and $\epsilon_T$ is the Trotter approximation
error, which scales as $\mathcal{O}\left(\frac{(2^m\tau)^3}{k^2}\right)$
\cite{Suzuki:92}. The Trotter approximation error can be made arbitrarily
small by increasing the integer Trotter parameter $k$. Since the
controlled-$U(2^m\tau)$ corresponds to the $(M-m)$th bit, $\epsilon_T$ must be
less than $1/2^{M-m}$, which is the precision of the $(M-m)$th measured bit in
the binary fraction for the phase $\phi$. Thus, when approximating
$U(2^m\tau)$, $k$ is increased until $\epsilon_T$ is less than $1/2^{M-m}$.
For a given $M$, we estimate a numerical value for the Trotter parameter
$k(m=0) = k_0$ as a function of $N \leq 10$, with the constraint that
$\epsilon_T < 1/2^M$. We thus find that for fixed $M$, $k_0$ scales as $1/N$.
We use the exponent based on $N \leq 10$ to extrapolate $k_0$ for larger $N$.
For $m>0$, we set $k=2^mk_0$, which will satisfy the error bound based on the
scaling of $\epsilon_T$ with $k$.

\begin{figure}
\centering
\setlength{\unitlength}{1mm}
\includegraphics[width=0.8\textwidth]{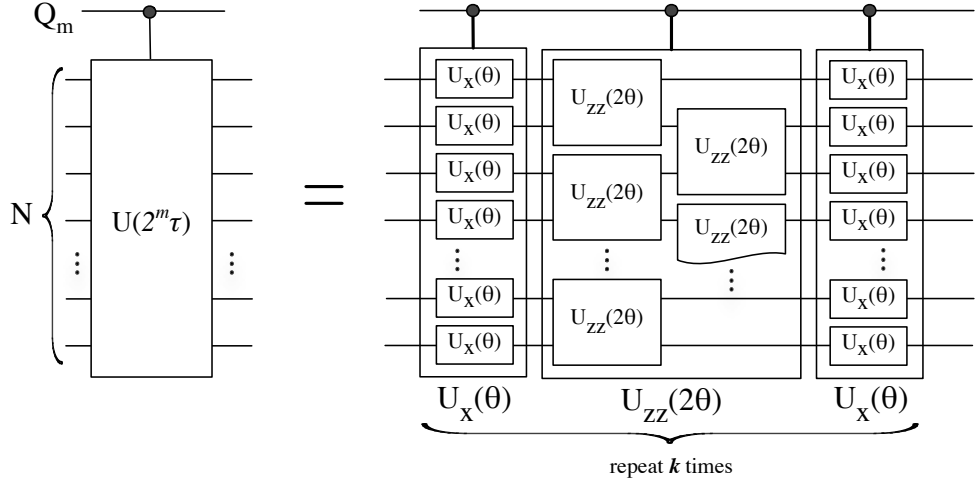}
\caption{Circuit for the controlled unitary operation $U(2^m\tau)$
approximated using the Trotter formula.}
\label{fig:UTrotter}
\end{figure}

The circuit corresponding to the Trotter approximation of
$U(2^m\tau)$ is shown in Figure \ref{fig:UTrotter}, where it can be seen
that the controlled-$U(2^m\tau)$ is composed of two
controlled-$U_x(\theta)$ operations and a controlled-$U_{zz}(\theta)$
operation, repeated $k$ times and controlled on the $m$th instance of the output
qubit denoted with $Q_m$. Expanding the circuit in Figure
\ref{fig:UTrotter}, we can express $\tilde{U}(2^m\tau)$ as:
\begin{equation}\label{eqn:UTrotter2}
\tilde{U}(2^m\tau) = U_x(\theta)\, \left [ 
U_{zz}(2\theta)U_x(2\theta) \right ]^{k-1}\, U_{zz}(2\theta) \, U_x(\theta),
\end{equation}
which shows that, approximating $U(2^m\tau)$ will require the sequential
implementation of $k$ controlled-$U_{zz}(2\theta)$ gates, $(k-1)$
controlled-$U_x(2\theta)$ gates, and two instances of controlled-$U_x(\theta)$
gates, all controlled on the $m$th instance of the output qubit.

\begin{figure}
\centering
\setlength{\unitlength}{1mm}
\includegraphics[width=1.0\textwidth]{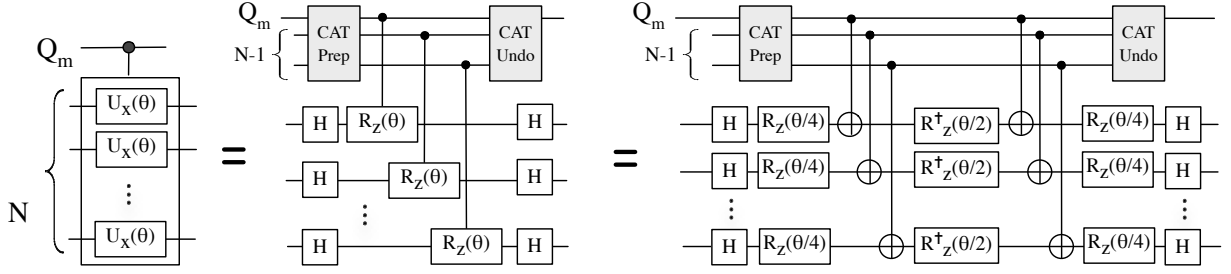}
\caption{The decomposition of the controlled unitary operation $U_x(\theta)$
into single-qubit $R_z$ gates and \cnot\ gates.}
\label{fig:controlled_Ux}
\end{figure}

\begin{figure}
\centering
\setlength{\unitlength}{1mm}
\includegraphics[width=1.0\textwidth]{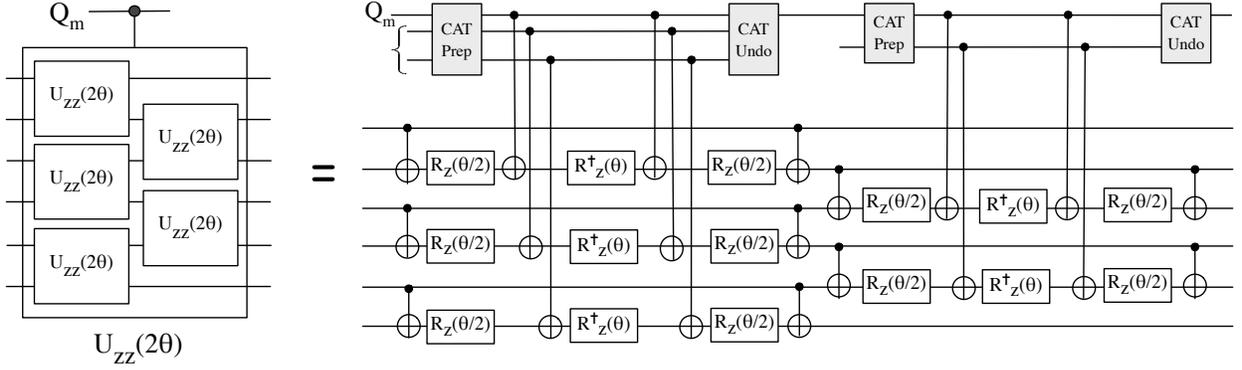}
\caption{The decomposition of the controlled unitary operation
$U_{zz}(2\theta)$ gate into single-qubit $R_z$ gates and \cnot\ gates.}
\label{fig:Uzz}
\end{figure}

The quantum circuits for the decomposition of the controlled-$U_x(2\theta)$ and
controlled-$U_{zz}(2\theta)$ gates are shown in Figures~\ref{fig:controlled_Ux} and
\ref{fig:Uzz}, respectively. The gates are decomposed into rotations about the $\hat
z$-axis, $R_z(\theta) = \exp(-i\frac{\theta}{2}Z )$ and \cnot\ gates.  $(N-1)$
additional qubits are used to prepare an $N$-qubit cat state in order to
parallelize each of the $N$ $R_z(\theta)$ gates. The preparation of an
$N$-qubit cat state requires $(N-1)$ \cnot\ gates, which can be implemented
in $\mathcal{O}(N)$ time steps in parallel with the $R_z(\theta/4)$ gates 
in Figure \ref{fig:controlled_Ux} and in parallel with the 
$R_z(\theta/2)$ gates in Figure \ref{fig:Uzz}.

The three single-qubit $R_z$ gates ($R_z(\theta)$, $R_z(\theta/2)$, and
$R_z(\theta/4)$) can be approximated using
$\mathcal{O}(\log^{3.97}(1/\epsilon_{sk}))$ basics gates ($H$, $T$,$S$) with
the Solovay-Kitaev theorem \cite{KSV,Dawson2005}. The Solovay-Kitaev error
($\epsilon_{sk}$) is equivalent to a small rotation applied to the qubit.
Using the results of Dawson and Nielsen \cite{Dawson2005} and 
$\theta = \frac{2^m\tau}{k}$,
to compute the sequence of $H$, $T$, and $S$ gates required to approximate each of
the three $R_z$ gates. We define $S_R$ as the length of the longest of
these three sequences.
For $M$=30, for example, we find that $S_R =
4\times10^5$, requiring a sixth order \cite{Dawson2005} Solovay-Kitaev
approximation. The results of this calculation show that the Solovay-Kitaev
error $\epsilon_{sk} < \displaystyle\frac{\epsilon_T}{k}$, in order that the
total error, $\epsilon_T$ is less than the required precision $(1/2^{M-m})$,
when we approximate $U(2^m\tau)$. As a result $S_R$ scales as
$\mathcal{O}(\log^{3.97}(k/\epsilon_T))=\mathcal{O}(M^{3.97})$. 

We now have a complete decomposition of the controlled-$U(2^m\tau)$ into the
basic gates given in Equation \ref{eqn:basicgates}. As a function of $S_R$,
the number of time steps required to implement controlled-$U_x(\theta)$ and
$U_{zz}(\theta)$ is equal to $(3S_R + 4)$, and $(6S_R + 7)$, respectively.
Following Equation \ref{eqn:UTrotter2}, the number of time steps required to
implement the entire controlled-$U(2^m\tau)$ is $k(9S_R + 11) + 3S_R + 4$,
where $k = 2^mk_0$. Each $R_j$ gate in Figure \ref{fig:OneControlQubit} is
equivalent to at most a rotation by $R_z(\theta)$ and requires less than $S_R$
gates.

Putting all of the above together, the total number of time steps ($K$)
required to implement the TIM circuit as a function of $S_R$, $k_0$, and $M$
is given by:
\begin{eqnarray}\label{eqn:TIMCost}
\nonumber K &=& \displaystyle \sum_{m=0}^{M-1} [2^mk_0(9S_R + 11) + 3S_R + 4 + S_R] \\
		     &=& \mathcal{O}(2^M)\times S_R,~~~ (M \to \infty) 
\end{eqnarray}
Since $S_R$ scales as $\mathcal{O}(M^{3.97})$, the total number of time steps 
is dominated by the exponential dependance on the precision (M). The
number of qubits $Q$ required to implement the circuit is $2N$, since $N$
qubits are needed for the input register $\ket{\Psi}$, one qubit is needed for
the output register, and $N-1$ qubits are needed for the cat state.

In the next section we include fault-tolerant QEC into our circuit model and
determine the resulting resource requirements, $K$ and $Q$. We also provide an
estimate on how long it could take to implement the TIM problem in real-time
by taking into account the underlying physical implementation of each gate and
qubit in the context of the QLA architecture.


%

\subsection{Mapping onto the QLA architecture\label{sec:faultTolerance}} 

Incorporating quantum error correction and fault-tolerance
\cite{Shor:1996ve,Aharonov97a,Kitaev97,Steane.A:1999aa} into the TIM
circuit design will impact the resource requirements in two ways.
First, each of the qubits becomes a {\em logical qubit}, that is
encoded into a state using a number of lower-level qubits. Second,
each gate becomes a {\em logical gate}, realized via a circuit
composed of lower-level gates applied on the lower-level qubits that
make-up a logical qubit. Each lower-level qubit may itself be a
logical qubit all the way down to the physical level. Thus, quantum
error correction and fault-tolerance increases the number of physical
time steps and qubits required to implement each basic gate and may
even require additional logical qubits, depending on how each gate is
implemented fault-tolerantly and the choice of error correcting code.
The resource requirements necessary to implement encoded logical
qubits and gates will depend on the performance parameters of the
underlying physical technology, the type of error correcting code
used, and the level of reliability required per logical operation. The
physical technology performance parameters that are taken into account
in the design of the QLA architecture are the physical gate
implementation reliability, time to execute a physical gate, and the
time it takes for the state of the physical qubits to decohere.

The QLA architecture \cite{Metodi:05} is a tile-based, homogeneous quantum
computer architecture based on ion trap technology, employing 2-D surface
electrode trap structures \cite{Cirac95,Wineland03,Chiaverini.J:2005aa}. Each
tile represents a single computational unit capable of storing two logical
qubits and executing fault-tolerantly any logical gate from the basic gate set
given in Equation \ref{eqn:basicgates}. One of the key features of the QLA
architecture is the teleportation-based logical interconnect which enables
logical qubit exchange between any two computational tiles. The interconnect
uses the entanglement-swapping protocol \cite{Dur.W:1999aa} to enable logical
qubit communication without adding any overhead to the number of time
required to implement a quantum circuit \cite{Metodi:05}.

The QLA was originally designed based on the requirement to factor $1024$-bit
integers  \cite{Metodi:05}. This requirement resulted in the need to employ
the second order concatenated Steane \steane quantum error correcting code
\cite{Steane.A:1996aa}. Second order concatenation means that each logical
qubit is a level $2$ qubit, composed of $7$ level $1$ logical qubits each
encoded into the state of $7$ physical ion-trap qubits. 


To estimate the reliability for executing each of the basic-gates
fault-tolerantly, a lower-bound of $3.1\times 10^{-6}$ for the fault-tolerant
threshold of the \steane code.  This value was derived by Metodi, et al
\cite{Metodi.T:2007aa}, by analysis of the ion-trap-based geometrical layout
of each logical qubit tile. The \steane code threshold value used in the
current research differs from the previously published estimate of $1.8\times
10^{-5}$ \cite{Svore06} due to our more detailed account of the operations
specific to the ion trap technology in the implementation of each logical
qubit \cite{Metodi.T:2007aa}. Gottesman's methodology
\cite{Gottesman.D:2000aa}, which takes into account qubit movement, and these
threshold results we estimate the reliability for each logical operation at
levels $1$ and $2$. 

Since each qubit in the \steane code moves an average of $10$ steps during
error correction \cite{Metodi.T:2007aa}, we find that each level $1$ gate has
a failure probability of $3.2\times 10^{-10}$ and each level $2$ gate has a
failure probability of $3.5\times 10^{-14}$. In our failure probability
estimates, we have assumed optimistic physical ion trap gate error
probabilities of $10^{-7}$ per physical operation, consistent with recent
ion-trap literature \cite{Ozeri.R:2005aa}. We also determine the physical
resources required for each logical qubit. Each level $1$ qubit requires $21$
ion-trap qubits ($7$ data qubits and $14$ ancilla to facilitate error
correction) and each level $2$ qubit requires $21$ level $1$ qubits. Given
that the duration of each physical operation on an ion-trap device is
currently on the order $10~\mu$s \cite{Seidelin.S:2006aa,Reichle06}, the time
required to complete a single error correction step is approximately $1.6$ ms
at level $1$ and $0.26$ seconds at level $2$.

\begin{figure}
\centering
\setlength{\unitlength}{1mm}
\includegraphics[width=0.6\textwidth]{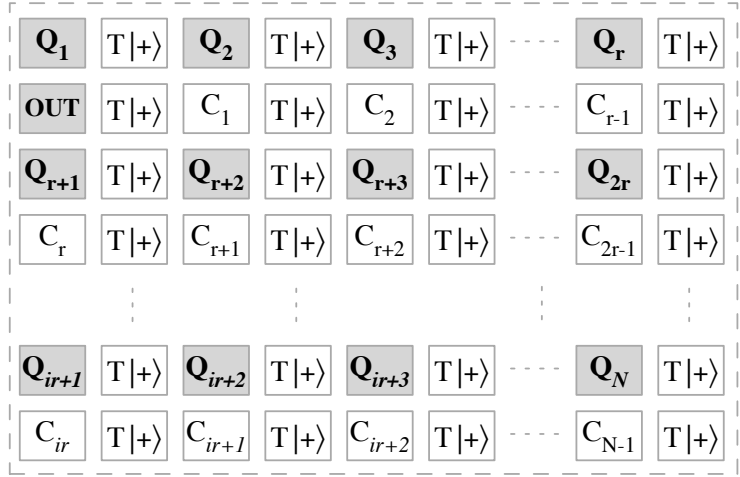}
\caption{QLA architecture for the TIM problem \label{fig:IsingArch}}
\end{figure}

The number of logical qubits $Q$ directly maps to the number of computational
tiles required by the QLA, allowing us to estimate the size of the physical
system. Similarly, the number of time steps $K$ maps directly to the time
required to implement the application since the duration of a single time step
in the QLA architecture is defined as the time required to perform error
correction, as discussed in Reference \cite{Metodi:05}. We define an
aggregated metric $KQ$ called the problem size equal to $K\times Q$, which is
an upper bound on the total number of logical gates executed during the
computation \cite{Steane:2002fj}. The inverse of the problem size, $1/KQ$, is
the maximum failure probability allowed in the execution of a logical gate
\cite{Steane:2002fj}, which ensures that the algorithm completes execution at
least $36$\% of the time. Taking into consideration the failure probabilities
per logical gate, the maximum problem size $KQ$ which can be implemented in
the QLA architecture is $3.1\times 10^9$ at level $1$ error correction,
$3\times 10^{13}$ at level $2$, and $2.8\times 10^{20}$ at level $3$. Level
$3$ error correction is not described in the design of the QLA architecture,
however, its implementation is possible since a level $3$ qubit is simply a
collection of level $2$ qubits and the architecture design does not change.
The estimated failure probability for each level $3$ logical gate is
$3.6\times 10^{-21}$.

The parameters $K$ and $Q$ for the TIM problem were estimated in Section
\ref{sec:dftg}, where $Q$ was found to be $2N$ and $K$ is on the order of
$\mathcal{O}(2^M)\times S_R$. The fault-tolerant implementation of the $T$
gate, however, requires an auxiliary logical qubit prepared into the state
$T\ket{+}$ for one time step followed by four time steps composed of $H$,
\cnot, $S$, and \meas~ gates \cite{Aliferis:2005kx}, causing the value of $K$
and $Q$ to increase. Since many of the gates in the Solovay-Kitaev sequences
approximating the $R_z$ gates are $T$ gates, when calculating $K$ using
Equation \ref{eqn:TIMCost}, the value of $S_R$ must take into consideration
the increased number of cycles for each $T$ gate. All other basic gates are
implemented transversally and require only one time step.

The resulting functional layout for the QLA architecture for the TIM
problem is shown in Figure \ref{fig:IsingArch}. The architecture consists of
$4N$ logical qubit tiles. The tiles labled with $Q_1$ through $Q_N$ are the
data tiles which hold the logical qubits used in the $N$-qubit input register
$\ket{\Psi}$ and the ``OUT'' tile is for the output register. The tiles labled with $C_1$
through $C_{N-1}$ are the $N-1$ qubit tiles for the cat state. The $T\ket{+}$
tiles are for the preparation of the auxiliary states in the event that $T$
gates are applied on any of the data qubits. All tiles are specifically
arranged as shown in Figure \ref{fig:IsingArch} in order to minimize the
communication required for each logical \cnot~gate between the control and
target qubits. For example, when preparing the cat state using all $C_i$ tiles
and the ``OUT'' tile, \cnot~gates are required only between the ``OUT'' tile,
$C_1$, and $C_r$. Similarly, $C_1$ interacts via a \cnot~gate only with $C_2$,
while $C_2$ interacts only with $Q_3$, during the cat state preparation.

\begin{figure}
\centering
\setlength{\unitlength}{1mm}
\includegraphics[width=1.0\textwidth]{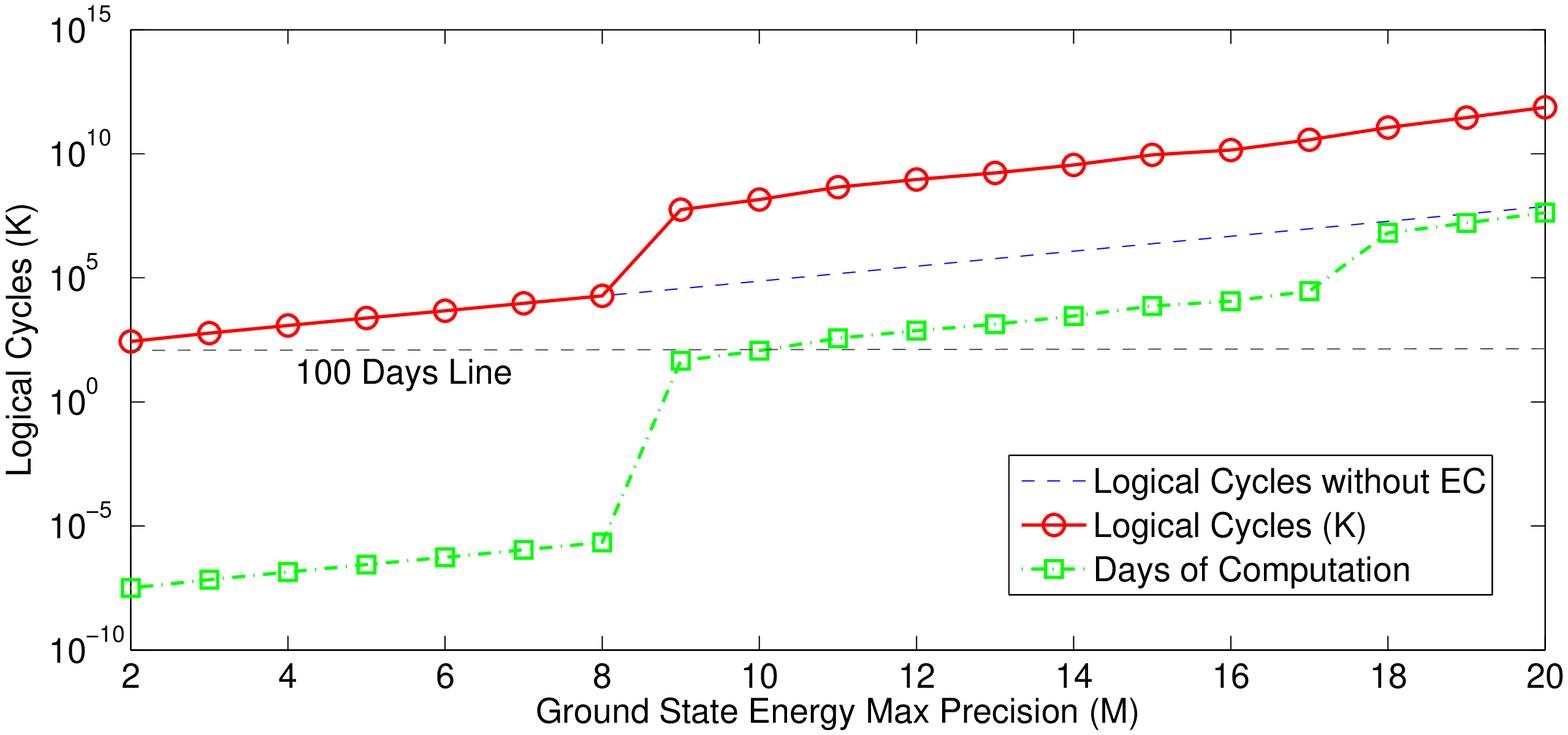}
\caption{(color online) Numerical calculations for the number of logical cycles $K$ (solid
line) and days of computation necessary assuming $N = 100$ spin TIM problem as a function of the
desired maximum precision $M \leq 20$.}\label{fig:Ising100_20}
\end{figure}

\subsection{Resource estimates for the 1-D TIM problem}\label{sec:oneD}

The resource requirements for implementing the 1-D TIM problem using the QLA
architecture are given in Figure \ref{fig:Ising100_20}, where we show a
logarithmic plot of the number of time steps $K$ (calculated using Equation
\ref{eqn:TIMCost}) as a function of the energy percision $M \leq 20$, assuming
$N = 100$. The figure clearly shows $K$'s exponential dependence on $M$. The
dependence of $K$ on the number of spins $(N)$ is negligible and appears only
in the $k_0$ term in Equation \ref{eqn:TIMCost} as $\mathcal{O}(1/N)$, as
discussed in Section \ref{sec:dftg}. In fact, since $Q = 4N$, we expect very
little increase in the value of the total problem size $KQ$ as $N$ increases.

We see that for $M \leq 8$ no error correction is required. This is because
the required reliability per gate of $1/KQ$ is still below the physical
ion-trap gate reliability of $1\times 10^{-7}$. Without error corection, the
architecture is composed entirely of physical qubits and all gates are
physical gates. This means that each single-qubit $R_z$ gate can be
implemented directly without the need to approximate it using the
Solvay-Kitaev theorem, resulting in $S_R = 1$ in Equation \ref{eqn:TIMCost},
and the total number of qubits becomes $2N$ instead of $4N$. For $M \geq 9$
error correction is required, resulting in a sudden jump in the number of
timesteps at $M = 9$, with an additional scaling factor of $\mathcal{O}(M^4)$
in $K$ due to $S_R$'s dependence on $M$. In fact, $K$ increases so quickly
that at $M = 9$ that level $2$ error correction is required instead of level
$1$. At $M \geq 18$ level $3$ error correction is required and while there is
no increase in $K$, each time step is much longer, so there is a jump in the
number of days of computation. The Solovay-Kitaev order \cite{Dawson2005} for
$M = 9$ is three and increases to order five for $M = 20$.

\subsection{Discussion of the resource estimates\label{sec:discussion}}

Our resource estimates for the 1-D TIM problem indicate that multiple levels
of error correction, even for modest precision requirements, results in long
computational times. As shown in Figure~\ref{fig:Ising100_20}, it takes longer
than $100$ days, even for $M = 7$, when level $2$ error correction is required.
When level $3$ error correction is required the estimated time is greater
than $7.5 \times 10^3$ {\em years}.

The number of logical cycles $K$, which grows exponentially with $M$,
contributes to the long computational times. However,  the primary factor
contributing to the long computational time is the time it takes to implement
a single logical gate using error correction. Presently, it is difficult to
see how one might reduce the value of $K$ short of implementing a different
approach for solving quantum simulation problems. On the other hand, the
logical gate time can be improved by implementing small changes in three
parameters: decreasing the physical gate time $t_p$, increasing the threshold
failure probability $p_{th}$, and decreasing the underlying physical failure
probability $p_0$. 

\begin{figure}
\centering
\setlength{\unitlength}{1mm}
\includegraphics[width=0.9\textwidth]{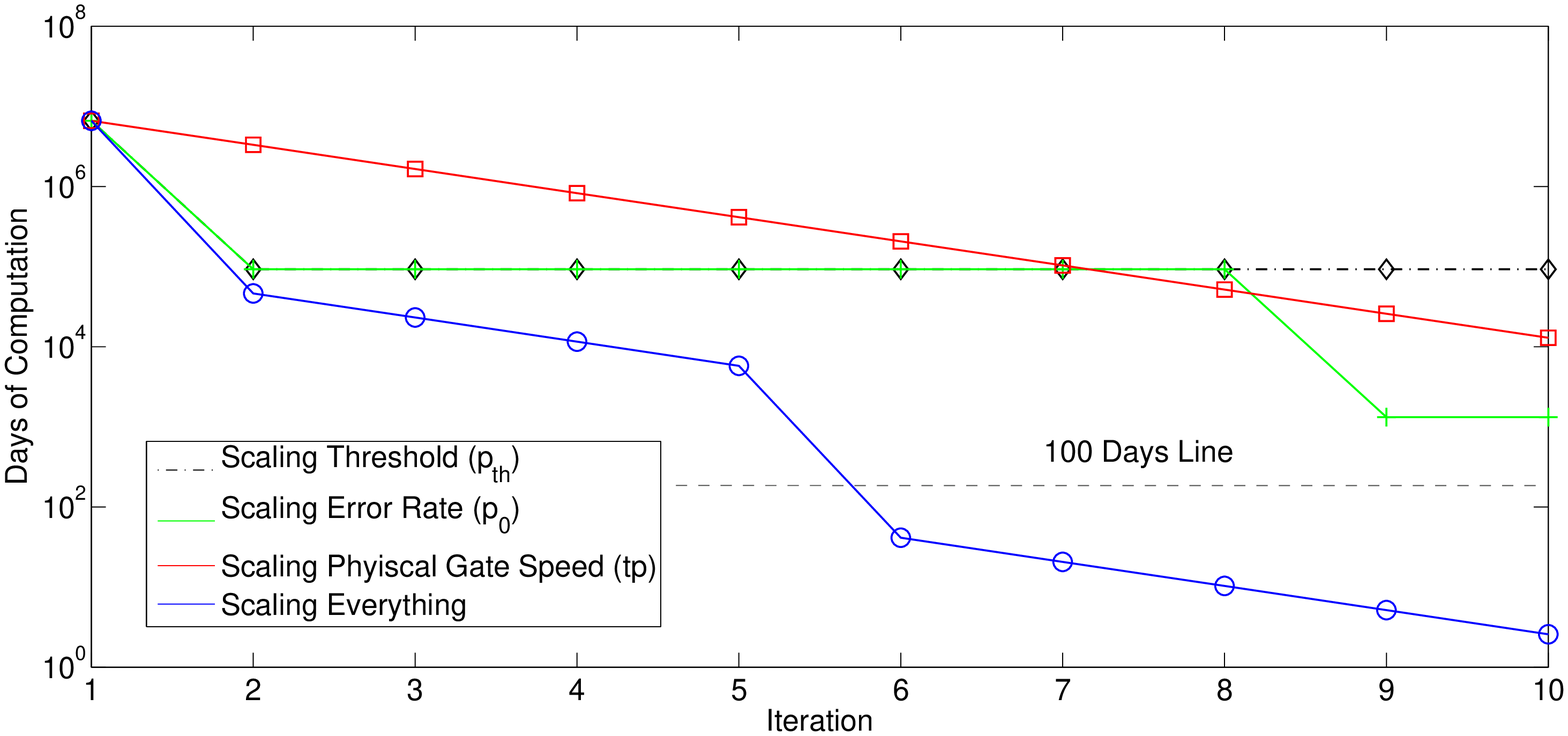}
\caption{(color online) The total computation time in days as we vary
the physical cycle time $t_p$ (square markers), physical failure probability
$p_0$ (starred markers), threshold failure probability $p_{th}$ (diamond
markers), and all together (circular markers) by a factor of two over 
10 iterations.\label{fig:days_scale}}
\end{figure}

The effect of these three parameters on the overall computational time for the
1-D TIM problem is shown in Figure \ref{fig:days_scale}. The figure shows how
the total time, in days, for $M=18$ varies as we improve each of the three
parameters by a factor of 2 during each of the $10$ iterations shown. The
starting values for each parameter in the figure are $3.1\times 10^{-6}$ for
$p_{th}$, $10^{-7}$ for $p_0$ and $10~\mu$s for $t_p$. 
Decreasing the physical failure probability and increasing the threshold
values by a factor of $2$ during each iteration causes the number of days to
decrease quadratically whenever lower error correction level is required,
otherwise the number of days remains constant from one iteration to the next.
A single change in the error correction level from level $3$ to level $2$
occurs by increasing $p_{th}$ by a factor of $2$  but there is no gain from
additional increases in the threshold alone. Decreasing only $p_0$ by a factor
of 512 yields two changes in the error correction level.


From this analysis, we see that in order to reach a computational time on the
order of $100$ days with only level $1$ error correction, we need
to achieve parameter values of $p_{th} = 1\times 10^{-4}$, $p_0 = 3
\times 10^{-9}$, and $t_p = 300$ ns, or better.
This provides goals for the improvement in the device technologies necessary
for quantum simulation. It should also be noted that these parameters are not
completely independent and improvements in one of them may result in improvements
in the others. For example, improving the physical failure
probability may lead to better threshold failure probability by
allowing some of the underlying operations to be weighted against one another.
Similarly improving the threshold failure probability, may require choosing a
more efficient quantum error correcting code which may have fundamentally
shorter logical time.

\section{Generalizing to higher spatial dimensions\label{sec:higherDim}}


The 1-D TIM ground state energy can be efficiently computed using classical
computing resources by taking advantage of the linear geometry of the spin
configuration and significantly reducing the effective state space to a
polynomial in $N$ \cite{Jouzapavicius1997}. A 2-D TIM with ferromagnetic and
antiferromagnetic Ising couplings can be difficult to solve due to spin
frustration. Many reductions to this problem still yield an exponential number
of states with near degenerate energy \cite{Moessner2003}. As a result, the
problem size scales exponentially with the size of the lattice. 
In contrast, the implementation of the quantum phase estimation circuit in
Figure~\ref{fig:OneControlQubit} is largely independent of the geometry of the
$N$ spin states and the values of $\Gamma_i$ and $J_{ij}$, which suggests
that it can be used for implementing efficiently higher-dimensional TIM problems.
Consider, for example, the calculation of the ground state energy
for the 2-D Villain's model  \cite{Moessner2001} using the phase estimation circuit.

Villain's model is a 2-D square lattice Ising model with $N^2$ spin sites
in which the rows have all ferromagnetic coupling and the
columns alternate between ferromagnetic and antiferromagnetic. 
Each of the $N^2$ sites in Villain's model are represented by
$N^2$ qubits in a $ N \times N$ grid. The only change to the circuit for the
phase estimation algorithm is the application of $U_{zz}$ Ising interaction,
which must be decomposed into two successive steps. First the rows of spin
states are treated as the 1-D TIM problem in parallel, followed by the
columns. Since the $U_{zz}$ operations within each step are done in parallel,
we still require $N/2$ additional qubits for the cat-states. Given that the
remaining operations, including the Quantum Fourier Transform implementation,
remain the same, the increase in the number of time steps to implement an
$N^2$-spin 2-D TIM problem, compared to the 1-D TIM problem, is by less than a
factor of two. Similarly, the increase in the resource requirements between a
1-D and a 3-D TIM problem will be by less than a factor of three. 


\section{Comparison with Factoring\label{sec:facCompare}}

Since the QLA architecture was initially evaluated in the context of Shor's
quantum factoring algorithm \cite{Shor:1994qy}, it would be interesting to
consider how the resource requirements for implementing the TIM problem
compare to those for implementing the factoring algorithm. In this section,
we compare the implementation of the two applications on the QLA architecture
and highlight some important differences between each application. 

Even though both applications employ the phase estimation algorithm, there are
several important differences. First, the precision requirements are
different. For Shor's quantum factoring algorithm, the precision $M$ must
scale linearly with the size $N$ of the $N$-bit number being factored
\cite{Shor:1994qy}, where $N \geq 1024$ for modern
cryptosystems. For quantum simulations, the desired precision is independent
of the system size N, and the required M is small compared to factoring. The
second difference lies in the implementation cost of the repeated powers of
the controlled-$U(\tau)$ gates for each application. In Shor's algorithm, the
gate is defined as $U(\tau)\ket{x}=\ket{ax~mod~N}$. Higher order powers of the
unitary can be generated efficiently via modular exponentiation
\cite{Shor:1994qy}. The result is that the implementation of $U(2^m\tau)$
requires $2m$ times the number of gates used for $U(\tau)$. For generic
quantum simulation problems, the implementation cost of $U(2^m\tau)$ equals
$2^{m}$ times $U(\tau)$, because of the Trotter parameter $k$. The
implementation of the control unitary gates for quantum simulation is not as
efficient as that for the modular exponentiation unitary gates. The third
difference lies in the preparation of the initial $N$-qubit state
$\ket{\Psi}$. The preparation of $\ket{\Psi}$ for the TIM problem by adiabatic
evolution is comparable in resource requirements to the phase estimation
circuit. For Shor's quantum factoring algorithm $\ket{\Psi} = \ket{1}$ in the
computational basis and is easily prepared.

\begin{figure}
\centering
\setlength{\unitlength}{1mm}
\includegraphics[width=1.0\textwidth]{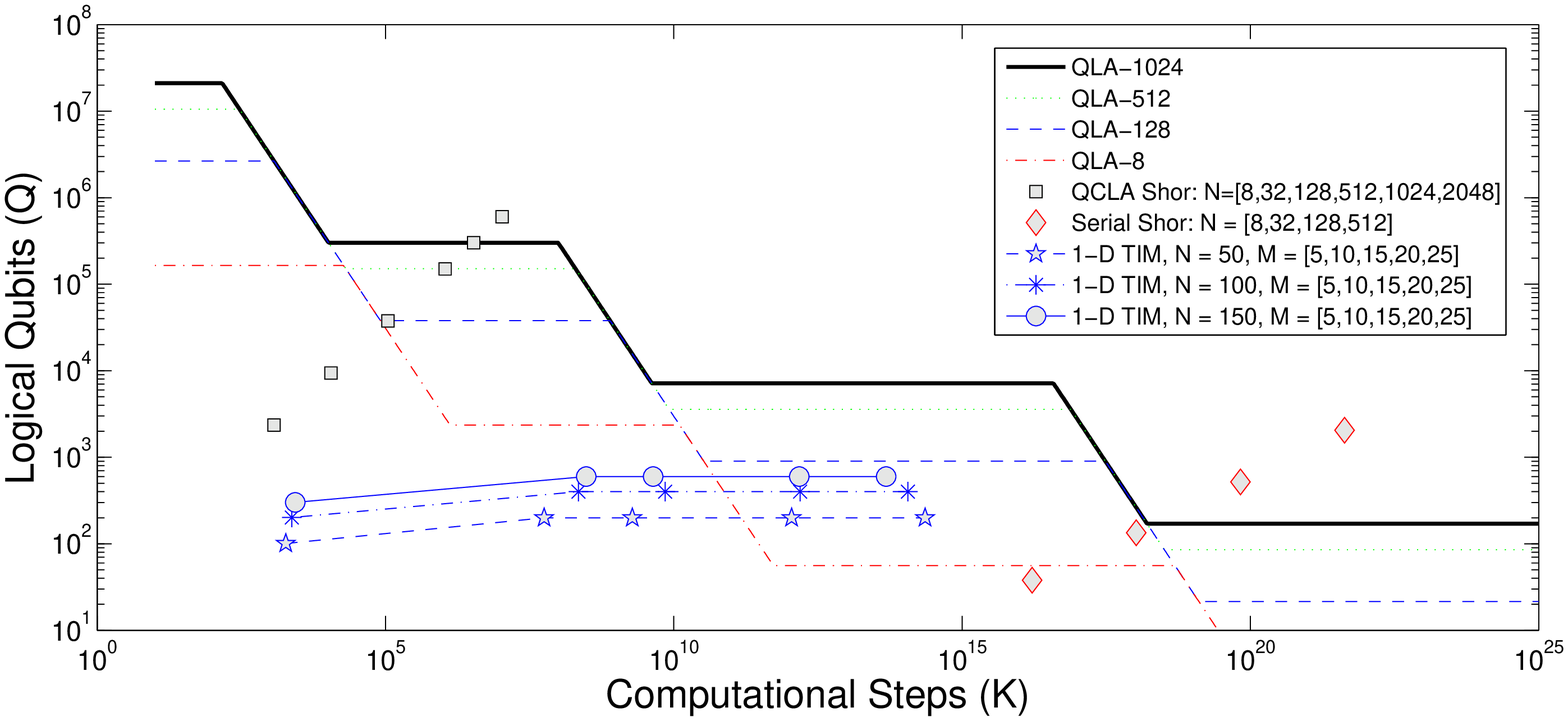}
\caption{(color online) Performance characteristics of different QLA-based 
quantum computers in KQ space with fixed amount of physical resources. 
The binary precision for the Ising problem of $M = \{5,10,15,20,25\}$ corresponds 
to decimal precision of $\{1,3,4,6,7\}$ digits, respectively. }\label{fig:testplot}
\end{figure}

Finally, factoring integers large enough to be relevant for modern
cryptanalysis requires several orders of magnitude more logical qubits
than the scale of quantum simulation problems considered in this
paper. At minimum, the factoring of an $N$-bit number requires $2N+3$
qubits, using the same one-control qubit circuit given in Figure
\ref{fig:OneControlQubit} \cite{Beauregard:2003qq}. As shown later in
this section, however, choosing to use only the minimum number of
qubits required for factoring leads to very high error correction
overhead. A more reasonable implementation of the factoring algorithm
requires $\mathcal{O}(N^2)$ number of logical qubits, which
corresponds to millions of logical qubits for factoring a $1024$-bit
number. Quantum simulation problems require significantly less
computational space and the problems considered in this paper require
less than $500$ logical qubits.


We examine how these differences affect the relative size of the QLA
architecture required to implement each application. In particular,
Figure \ref{fig:testplot} shows the performance of QLA-based quantum
computers in KQ space with fixed physical resources. Each horizontal
line corresponds to the KQ limit for a QLA-based architecture modeled
for factoring a $1024$-bit number (top-most horizontal dashed line), a
$512$-bit number, a $128$-bit number, and an $8$-bit number,
respectively. The physical resources for each QLA-N quantum computer
(where $N=\{1024,512,128,8\}$ bits) are determined by how many logical
qubits at level $2$ error correction are required to implement the
Quantum Carry Look-ahead Adder (QCLA) factoring circuit
\cite{Meter.R:2005aa,Metodi:05}, which requires $\mathcal{O}(N^2)$
logical qubits and $\mathcal{O}(N~log^2N)$ logical cycles. The
plateaus in each QLA-$N$ line of Figure \ref{fig:testplot} represent
using all of the qubits at a specific level of encoding, with the
top-most right-hand plateau representing level $1$. Where the lines
are sloped, the model is that only a certain number of the lower level
encoded qubits can be used. Once this reaches the number of qubits
that can be encoded at the next level, the quantum computer is
switched from encoding level $L$ to $L+1$ by using all the available
level $L$ qubits.

Figure \ref{fig:testplot} shows that a QLA-$N$ quantum computer is
capable of executing an application using level $L$ encoded qubits if
the application instance is mapped {\em underneath} the line
representing the computer at level $L$. Factoring a $1024$-bit number,
for example, falls directly on the level $2$ portion of the QLA-$1024$
line (see the square markers). Anything above that line cannot be
implemented with the QLA-$1024$ computer. Similarly, factoring a
$128$-bit number maps under the QLA-$128$ line, but can be
accomplished using level $1$ qubits. The TIM problem is mapped onto
Figure \ref{fig:testplot} for $N=50$, $100$, $150$ and several binary
precision instances: $M = \{5,10,15,20,25\}$. As expected, factoring
requires many more logical qubits, however, both applications require
similar levels of error correction. A decimal precision of up to $4$
digits of accuracy ($M = 15$) can be reached by using a quantum
computer capable of factoring an $8$-bit number at level $2$ error
correction, however higher precision quickly requires level $3$ error
correction.

The resources for implementing quantum factoring with
one-control-qubit were calculated following the circuit in Figure
\ref{fig:OneControlQubit}, where the unitary gates are replaced with the
unitary gates corresponding to modular exponentiation, as
discussed in Reference \cite{Beauregard:2003qq}. The results are shown with the
diamond-shaped markers in Figure \ref{fig:testplot}. While this
particular implementation is the least expensive factoring network in
terms of logical qubits, the high precision
requirement of $M = O(N)$ makes this network very expensive in terms
of time steps. In fact, the number of time steps required pushes the
reliability requirements into level $4$ error correction for factoring
even modestly-sized numbers.

\section{Conclusion \label{sec:conclusion}}

In this paper, the TIM quantum simulation circuit was decomposed into
fault-tolerant operations and we estimated the circuit's resource
requirements and number of logical cycles $K$ as a function of the
desired precision $M$ in the estimate of the ground state energy. Our
resource estimates were based on the QLA architecture and underlying
technology parameters of trapped ions allowing us to estimate both
$K$, as a function of the level of the error correction level, and the
total length of the computation in real-time.

Our results indicate that even for small precision requirements $K$ is
large enough to require error correction. The growth of $K$ is due to
its linear dependence on the the Trotter parameter $k$, which scales
exponentially with the maximum desired precision $M$. In order for $K$
to scale polynomially with the precision, new quantum simulation
algorithms are required or systems must be chosen where the phase
estimation algorithm can be implemented without the Trotter formula.
The linear dependence of the number of time steps on $k$ is due to the
fact that $U_x$ and the $U_{zz}$ do not commute. However, there are
some physical systems, whose Hamiltonians are composed of commuting
terms, such as the nontransversal classical Ising model, which has a
solution to the partition function in two dimensions but is
NP-Complete for higher dimensions\cite{Istrail2000}. In those cases,
Trotterization is unnecessary. In future work, we intend to generalize
the calculations of the resource requirements to other physical
systems and consider different ways to implement the phase estimation
algorithm that limit its dependence on the Trotter formula.


\end{document}